\documentclass[11pt]{article}
\usepackage{geometry}                
\usepackage{graphicx}
\usepackage{amssymb}
\usepackage{amsmath}
\usepackage{epstopdf}
\usepackage{color}
\usepackage{soul}

\newcommand{\eps}{\varepsilon}
\newcommand{\R}{\mathbb{R}}
\newcommand{\Tr}{\mbox{Tr}}

\RequirePackage{color}
\newcommand{\red}[1]{{\color{black}{#1}}}
\newcommand{\blue}[1]{{\color{black}{#1}}}

 \definecolor{MyDarkGreen}{rgb}{0.02,0.60,0.06}

\title{Calibration with confidence: A  principled method for panel assessment}
\author{R.S.~MacKay$^1$, R.~Kenna$^2$, R.J.~Low$^2$  \& S.~Parker$^1$\\
$^1$ Mathematics Institute and Centre for Complexity Science\\
University of Warwick, Coventry CV4 7AL, U.K.\\[1ex]
$^2$ Applied Mathematics Research Centre\\
Coventry University, Coventry CV1 5FB, U.K.}
\date{08 February 2017}         

\begin{document}
\maketitle

\blue{In memory of Professor Sir David John Cameron MacKay (22 April 1967 - 14 April 2016)}

\begin{abstract}
Frequently, a set of objects has to be evaluated by a panel of assessors, but not every object is assessed by every assessor.
A problem facing such panels is how to take into account different standards amongst panel members and varying levels of confidence in their scores. 
Here, a mathematically-based algorithm is developed to calibrate the scores of such assessors, addressing both of these issues.
The algorithm is based on the connectivity of the graph of assessors and objects evaluated, incorporating declared confidences as weights on its edges. 
If the graph is sufficiently well connected, relative standards can be inferred by comparing how assessors rate objects they assess in common, weighted by  the levels of confidence of each assessment.
By removing these biases, ``true" values are inferred for all the objects.
Reliability estimates for the resulting values are obtained.
The algorithm is tested in two case studies, one by computer simulation and another based on realistic evaluation data.
The  process is compared to the simple averaging procedure in widespread use, and to Fisher's additive incomplete block analysis.
It is anticipated that the algorithm will
prove useful in a wide variety of situations such as evaluation of the quality of research submitted to national assessment exercises;  
appraisal of grant proposals submitted to funding panels; 
ranking of job applicants; and judgement of performances on degree courses wherein candidates can choose from lists of options.
\end{abstract}

\vspace{1cm}

{Keywords: {Calibration, evaluation, assessment, confidence, uncertainty, model comparison.}}

\newpage

\section{Introduction}

\blue{This paper} address\blue{es} the widespread problem of how to take into account differences in standards, confidence and bias in assessment panels, such as those evaluating research quality or grant proposals, employment or promotion applications\blue{,} and classification of university degree courses, in situations where it is not feasible for every assessor to evaluate every object to be assessed.

A common approach to assessment of a range of objects by such a panel is to assign to each object the average of the scores awarded by the assessors who evaluate that object.
This approach is represented by the {cell labelled ``simple averaging'' (SA)} in the top left of a matrix of approaches listed in Table~1, 
but it ignores the likely possibility that different assessors have different levels of stringency, expertise and  bias \cite{Meadows2006}.
Some panels shift the scores for each assessor to make the average of each take a normalised value, but this ignores 
the possibility that the set of objects assigned to one assessor may be of a genuinely different standard from that assigned to another.
For an experimental scientist, the issue is obvious: {\emph{calibration}}.

One  \blue{solution} is to seek to calibrate the assessors beforehand on a common subset of objects, perhaps disjoint from the set to be evaluated  \cite{Paul1981}. 
This means that they each evaluate all the objects in the subset and then some  rescaling is agreed to bring the assessors into line as far as possible.  
This would not work well, however, in a situation where the range of objects is broader than the expertise of a single assessor. 
Also, regardless of how well the assessors are trained, differences between individuals' assessments of objects remain in such ad hoc approaches \cite{Naes2010}.

\begin{table}[b!]
\begin{center}
  \begin{tabular}{ | l || l | l |}
    \hline
                      &  {\bf{Without }}  & {\bf{With }} \\ 
		                  & {\bf{confidences}} & {\bf{confidences}} \\ \hline \hline									
  {\bf{Without  }}   &  Simple  & Confidence-weighted  \\ 
	{\bf{calibration }}&  averaging (SA) & averaging \blue{(CWA)} \\ \hline
  {\bf{With}}&  {{Incomplete block }}   & Calibration with    \\
	{\bf{calibration }}&  {{analysis}} {{(IBA)}}  & confidence  {{(CWC)}} \\
    \hline
  \end{tabular}
\end{center}
\caption{Panel Assessment Methods: The matrix of four approaches according to use of calibration and/or confidences.  Simple averaging (SA) is the base for comparisons.
Fisher's IBA does not deal with varying degrees of confidence and the confidence-weighted averaging doesn't achieve calibration. 
The method proposed herein (CWC) accommodates both calibration and confidences.}
\end{table}

\begin{figure}[t!]
   \centering
\includegraphics[width=0.7\textwidth]{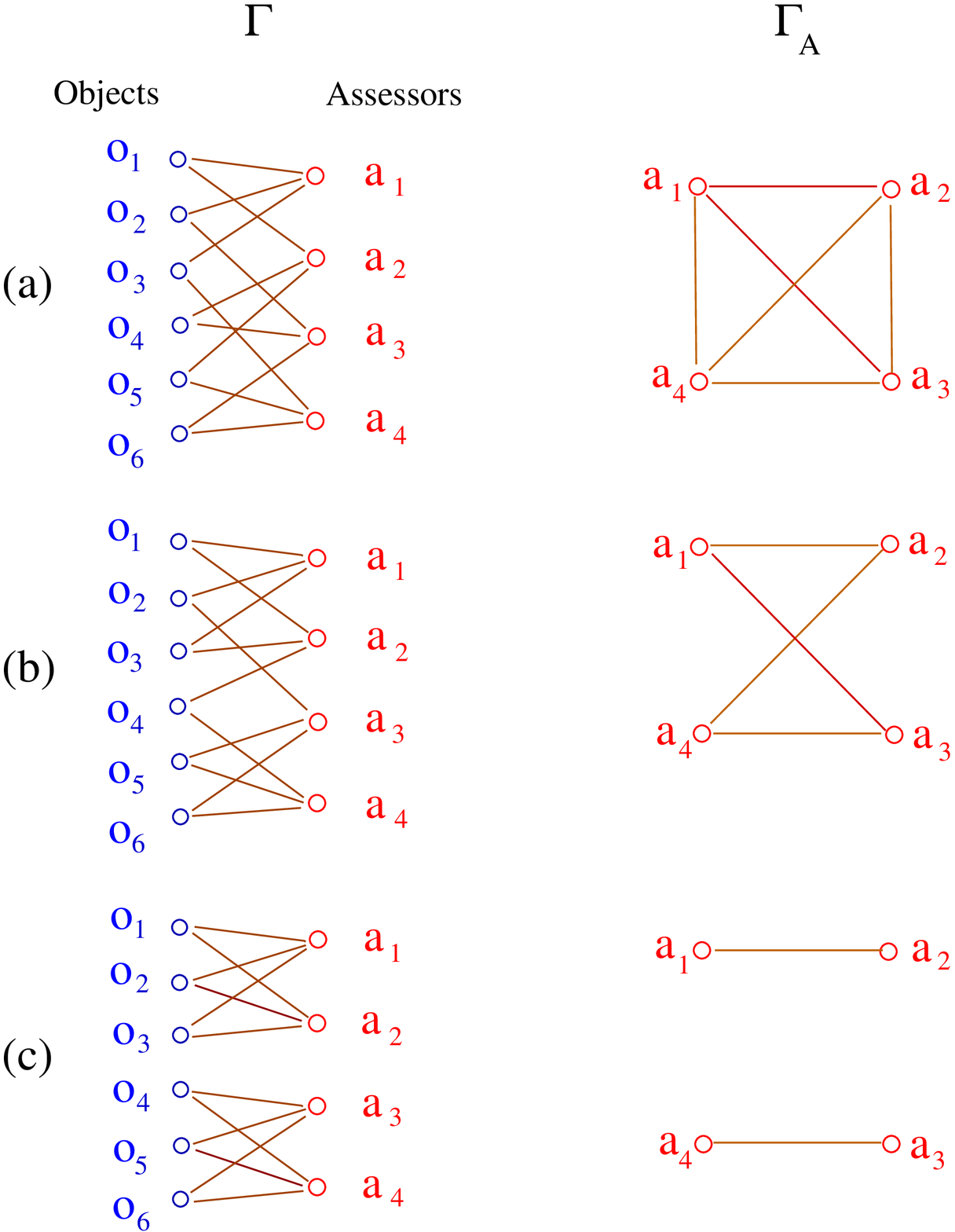}
 \vspace{0cm}
\caption{Three examples of assessment graphs $\Gamma$ showing which object $o_j$ is assessed by which assessor $a_k$, and the resulting graphs $\Gamma_A$ on the set of assessors where two assessors are linked if they assess an object in common.  Case (a) produces a fully connected assessor graph, (b) a moderately connected graph, whereas case (c) is disconnected.
 }
\label{fig1:graphs}
\end{figure}

If the expertise of two assessors overlap on some subject, however, any discrepancy between their evaluations can be used to infer information about their relative standards.  
Thus if the graph $\Gamma_A$ on the set of assessors, formed by linking two whenever they assess a common object, is sufficiently well connected  one can expect to be able to infer a robust calibration of the assessors and hence robust  scores for the  \blue{objects}.
The construction of this graph is illustrated in Figure~\ref{fig1:graphs}\blue{, beginning from the graph $\Gamma$ showing which objects are assessed by which assessors}.  

One approach to achieving such calibration was developed by  R.A.Fisher \cite{Fish}, in the context of trials of crop treatments.
Denoting the score from assessor $a$ for object $o$ by $s_{ao}$, Fisher's approach is based on fitting a model of the form $s_{ao} = v_o+b_a+\eps_{ao}$ with $\eps_{ao}$ independent identically distributed random variables of mean zero.  Then $b_a$ is the bias inferred for assessor $a$ and $v_o$ is the value inferred for object $o$.
Fisher's approach is known as {\em additive incomplete block analysis}  (IBA) and a  body of associated literature 
and applications has since been developed \cite{G}, though its use in panel assessment seems rare. 
It is  represented as the bottom left entry of Table~1. 

Another ingredient that is important in many panel assessments, however, is  different weights that may be put on different assessments.  
We refer to these weights as ``confidences".  
Fisher's IBA  does not  take different levels of confidence into account.

If the assessors express confidences in the assessments, for example by some pre-determined weights assigned to types of assessment or by the assessors declaring confidences in each of their scores, then it is natural to replace simple averaging by confidence-weighted averaging (CWA).
This is represented as  the top right element of Table~1, but 
 it doesn't address the calibration issue so we do not consider it further.

In this paper we present and test a method to calibrate scores taking into account confidences, that is, we complete the bottom-right corner of the matrix of approaches represented in Table~1, where our method is termed {\em calibration with confidence} (CWC).
We demonstrate that the method can achieve a greater degree of accuracy with fewer assessors than  either simple averaging or IBA, and we derive robustness estimates taking the confidences into account.

We  are aware of two other schemes that incorporate confidences into a calibration process.
One is the abstract-review method for the SIGKDD'09 conference (section 4 of \cite{Fl}; see also \cite{Gu}).  
The other is the abstract-review method used for the NIPS2013 conference (building on \cite{PB} and described in \cite{GWG}).  
Our method has the advantages of simplicity of implementation and  a straightforward robustness analysis.
We leave detailed comparison  with methods such as these for future publication.

\section{The model}
\label{sec:model}

Let us  suppose that each assessor is assigned a subset of the objects to evaluate.  
Denote the resulting set of (assessor, object) pairs by $E$.
Let us further suppose that the score $s_{ao}$ that assessor $a$ assigns to object $o$ is a real number related to a ``true" value $v_o$ for the object by
\begin{equation}
s_{ao} = v_o + b_a + \eps_{ao},
\label{basic}
\end{equation}
where $b_a$ can be called the {\em bias} of assessor $a$ and $\eps_{ao}$ are independent zero-mean random variables.
Such a model forms the basis for additive incomplete block analysis. This was also proposed in ref.~\cite{TDF} (see equation~(8.2b) therein) but without a method to  estimate the true  values.  
Here we will achieve this and make a significant improvement, namely the incorporation of varying confidences in the scores.

To take into account the varying expertise of the assessors\blue{ with respect to the objects}, we propose that in addition to the score $s_{ao}$, each assessor is asked to specify a level of confidence for that evaluation.  
This could be in the form of a rating such as  ``high'', ``medium'', ``low'', as requested by some funding agencies, but we propose \blue{to allow} something more general and akin to experimental science.
Confidence can be estimated by asking assessors to specify an uncertainty $\sigma_{ao} > 0$ for their score and then the confidence level (or ``precision'') is taken to be
\begin{equation}
c_{ao} = 1/\sigma_{ao}^2.
\label{eq:conf}
\end{equation}
The instructions to the assessors can be  \blue{to} choose $s_{ao}$ and $\sigma_{ao}$ so that $2/3$ of their probability distribution for the score lies in $[s_{ao}-\sigma_{ao}, s_{ao}+\sigma_{ao}]$, $1/6$ above this interval and $1/6$ below it. Methods for training assessors to estimate uncertainties are presented in \cite{H}.
\blue{There are also methods for training assessors on the assessment criteria to improve their accuracy \cite{S+}, which could also be expected to have the beneficial effect of reducing their uncertainties.}

So let us suppose that
\begin{equation}
\eps_{ao} = \sigma_{ao} \eta_{ao},
\label{etaw}
\end{equation}
with $\eta_{ao}$ independent zero-mean, random variables of 
common variance $w$. 
For the moment, we set $w=1$; extensions to other values of $w$ are considered in Appendix A, and in particular are necessary if confidence is expressed only qualitatively.
\blue{In the case that confidences are reported as only high, medium or low, they can be converted into quantitative ones by for example choosing $\lambda \approx 2$ and setting $c_{ao} = \lambda^2, 1, \lambda^{-2}$, respectively.  
The interpretation of $\lambda$ is the ratio of the uncertainty for a low confidence evaluation to that for a medium one, and for a medium one to a high one.
Then $w$ is unspecified but can be fit from the data, as in Appendix A.}

Thus our basic model is
\begin{equation}
s_{ao} = v_o + b_a + \sigma_{ao} \eta_{ao}.
\label{eq:model1}
\end{equation}

\section{Solution of the model}
\label{sec:solution}

Given the data $\{(s_{ao}, \sigma_{ao}) : (a,o) \in E\}$ for all assigned assessor-object pairs, we wish to extract the true values $v_o$ and assessor biases $b_a$.  The simplest procedure is to minimise the sum of squares 
\begin{equation}
\sum_{(a,o) \in E} \eta_{ao}^2 =   \sum_{(a,o)\in E} c_{ao} (s_{ao} - v_o - b_a)^2 ,
\label{sumsquares}
\end{equation}
where the confidence level $c_{ao}$  was defined in Equation (\ref{eq:conf}). 
This procedure can be justified if the $\eta_{ao}$ are assumed to be normally distributed, because then it gives the maximum-likelihood values for $v_o$ and $b_a$.  It can also be viewed as orthogonal projection of the vector $s$ of scores $s_{ao}$ to the subspace of the form $s_{ao}=v_o+b_a$ in the \blue{Riemannian} metric given by $|s| = \sqrt{\sum_{ao} c_{ao}s_{ao}^2}$.

Now expression~(\ref{sumsquares}) is minimised with respect to $v_o$ iff
$${\sum_{a:(a,o)\in E} c_{ao}(s_{ao}-v_o-b_a) = 0},$$
and with respect to $b_a$ iff
$$\sum_{o : (a,o) \in E} c_{ao} (s_{ao} - v_o - b_a) = 0. $$
It is notationally convenient to extend the sums to all assessors (respectively objects) by assigning the value $c_{ao} = 0$ to any assessor-object pair that is not in $E$ (i.e.~for which a score was not returned).  
Then these conditions can be written as
\begin{eqnarray}
C_o v_o + \sum_a b_a c_{ao} & = & V_o 
 \label{eq:system1}\\
\sum_o c_{ao} v_o + C^\prime_a b_a & = & B_a . 
\label{eq:system1b} 
\end{eqnarray}
Here, 
\begin{equation}
 V_o = \sum_{a} c_{ao} s_{ao}
 \label{Referee1a}
\end{equation}
is the confidence-weighted total score for object $o$ and 
\begin{equation}
 B_a = \sum_o c_{ao}s_{ao}
\label{Referee1b}
\end{equation}
 is that for assessor $a$,
\begin{equation}
 C_o = \sum_a c_{ao}
\label{eq:vob}
\end{equation}
is the total confidence in the assessment of object $o$ and
\begin{equation}
 C^\prime_a=\sum_o c_{ao}
 \label{eq:voc}
\end{equation}
is the total confidence expressed by assessor $a$.

Equations~(\ref{eq:system1}) and~(\ref{eq:system1b}) form a linear system of equations for the $v_o$ and $b_a$.
It has an obvious degeneracy in that one could add a constant $k$ to all the $v_o$ and subtract $k$ from all the $b_a$ and obtain another solution.  
 \blue{One} can remove this degeneracy by, for example, imposing the condition 
\begin{equation}
 \sum_a b_a = 0.
\label{bias}
\end{equation} 
This is the simplest possibility and corresponds to a translation \blue{(shift)} that brings the average bias over assessors to zero. 
Alternatives are discussed in Appendix B.

Define a graph $\Gamma$ linking assessor $a$ to object $o$ if and only if $(a,o) \in E$, as illustrated in the left column of Figure~\ref{fig1:graphs}.  The edges in the graph are weighted by the confidences $c_{ao}$.
Whether the set of equations~(\ref{eq:system1}) and (\ref{eq:system1b}) has a unique solution after breaking the degeneracy depends on the connectivity of  $\Gamma$.
Define a linear operator $L$ by writing equations~(\ref{eq:system1}) and (\ref{eq:system1b}) as
\begin{equation}
L \left[ \begin{array}{c} v \\ b \end{array} \right] = \left[ \begin{array}{c} V \\ B \end{array} \right],
\label{eq:L}
\end{equation}
where $v,b,V$ and $B$ denote the column vectors formed by the $v_o,b_a,V_o$ and $B_a$ respectively.
The operator $L$
has null space of dimension equal to the number of connected components of $\Gamma$ (this follows from Perron-Frobenius theory, see e.g.~ref.\cite{M}).  
Thus if $\Gamma$ is connected, the null space of $L$ has dimension one, so corresponds precisely to the null vectors $v_o = k\ \forall o, b_a =-k\ \forall a$, that we already noticed and dealt with.  Connectedness of $\Gamma$ ensures that if (\ref{eq:L}) has a solution then there is a unique one satisfying (\ref{bias}).

It remains to check that the right-hand side of equation~(\ref{eq:L}) lies in the range of $L$, thus ensuring that a solution exists.  This is true if all null forms of the adjoint operator $L^\dagger$ send the right-hand side to zero.  The null space of $L^\dagger$ has the same dimension as that of $L$, because $L$ is square, and an obvious non-zero null form \blue{$\alpha$} is given by
\begin{equation}
\alpha(v,b) = \sum_o v_o - \sum_a b_a .
\label{eq:alpha}
\end{equation}
It follows from the definitions of $V$ and $B$ that $\alpha(V,B) = 0$.  So a solution exists.

Thus under the assumption that the assessor-object graph $\Gamma$ is connected, equations (\ref{eq:system1}) and (\ref{eq:system1b})  have a unique solution $(v,b)$ satisfying equation~(\ref{bias}).  Note that  connectedness of $\Gamma$ is necessary for uniqueness, otherwise one could follow an analogous procedure, adding and subtracting constants independently in each connected component of $\Gamma$, and thereby produce more solutions.

The equations (\ref{eq:system1},\ref{eq:system1b}) have a special structure, due to the bipartite nature of $\Gamma$, that  can be worth exploiting.
The first equation~(\ref{eq:system1})  can be written as
\begin{equation}
v_o = \frac{V_o - \sum_a b_a c_{ao}}{C_o} .
\label{eq:vo}
\end{equation}
This can be substituted into the second equation~(\ref{eq:system1b}) to obtain 
\begin{equation}
\sum_{a'} C_{aa'} b_{a'} -  C^\prime_a b_a = \sum_o \frac{c_{ao}V_o}{C_o} - B_a ,
\label{eq:reduced}
\end{equation}
where
\begin{equation}
C_{aa'} = \sum_o \frac{c_{ao}c_{a'o}}{C_o}
\label{eq:Aweights}
\end{equation}
can be considered as weights on the edges of the graph $\Gamma_A$ on assessors illustrated in the right column of Figure~\ref{fig1:graphs}.
The dimension of the \blue{reduced} system (\ref{eq:reduced}) is the number $N_A$ of assessors (rather than the sum of the numbers of assessors and objects), which gives some computational savings.  
Replacing one of the equations in (\ref{eq:reduced}), say that for the ``last'' assessor, by equation~(\ref{bias}) gives a system with a unique solution that can be solved for $b$ by any method of numerical linear algebra, e.g.~LUP decomposition \cite{LUPref}.  Then $v$ can be  obtained from equation~(\ref{eq:vo}).

\blue{A slightly more sophisticated approach to incorporating a degeneracy-breaking condition into the equations (\ref{eq:reduced}) is described in Appendix B.}

A key question with any black-box solution like the one presented here is how robust is the outcome?  We propose two ways of quantifying the robustness.  One is to bound how much the outcome would change if some of the scores were changed (e.g.~representing mistakes or anomalous judgements).  We treat this in Appendix C.  
The other is to evaluate the posterior uncertainty of the outcomes, assuming normal distribution of the $\eta_{ao}$.  This is treated in Appendix D.

\section{Case Studies}

We have tested the approach in three contexts. 
We report in detail on two case studies here.
In the first case study, we use a computer-generated set of data containing true values of assessed items,  assessor biases and confidences  for the assessments, and resulting scores. 
This has the advantage of allowing us to compare the values obtained by the new approach with the true underlying value of each item. 
The second  case study is an evaluation of grant proposals  using realistic data based on a university's internal competition.
In this test, of course, there is no possibility to access ``true'' values, so instead we compare the evidence for the models using a Bayesian approach (Appendix E), and we compare their posterior uncertainties (Appendix D).
The third context in which we tested our method was assessment of students; we report briefly on this at the end of the section.

\subsection{Case Study 1 -- Simulation}
In the simulation, $N_O=3000$ objects are assessed by a panel of $N_A=15$ assessors.
This choice was motivated  by the number of outputs and reviewers in the applied mathematics unit of assessment at the UK's 2008 research assessment exercise.
The simulation was carried out using {\tt{MATLAB}}, and the system of equations was solved using its built-in procedure, which computed the LU decomposition of $L$ (with the last row replaced by the degeneracy-breaking condition (\ref{bias})).
The reduction to (\ref{eq:reduced}) was not used because $N_O = 3000$ is easily handled by modern personal computers.

True values of the items $v_o$ were  assumed to be normally distributed with a mean of $50$ and standard deviation set to $15$, but with $v_o$ values truncated at $0$ and $100$.
The assessor biases $b_a$ were  assumed to be normally distributed with a mean of $0$ and a standard deviation of $15$. 
Each assessment was considered to be done with  high, medium, or low confidence, and these were modelled using scaled uncertainties for the awarded scores, of $\sigma_{ao} = 5$, 10 or 15 respectively.
The allocated scores follow equation~(\ref{eq:model1}), but truncated at 0 and 100.

With $r$ assessors per item (which we took to be the same for each item in this instance), each simulation generated $rN_O$ object scores $s_{ao}$.
From these, we generated $N_O$ value estimates $\hat{v}_o$ and $N_A$ estimates of assessor biases $\hat{b}_a$ using the calibration processes.
We then took the mean and maximum values of the errors in the estimates,  $dv_o = |\hat{v}_o - v_o|$ and $db_a = |\hat{b}_a - b_a|$.
Simple averaging also delivered a value estimate $\hat{v}_o$, as well as mean and maximal values of the errors $dv_o$.
Finally, we determined the averages of the errors $dv_o$ and $db_a$ over 100 simulations. 
The results for these averaged mean and maximal errors in the scores are denoted by $\langle{dv}\rangle$ and $(dv)_{\rm{max}}$, respectively and those for the biases 
(for the calibrated approaches only) are denoted $\langle{db}\rangle$ and $(db)_{\rm{max}}$.

\begin{figure}[t!]
\includegraphics[width=0.5\textwidth]{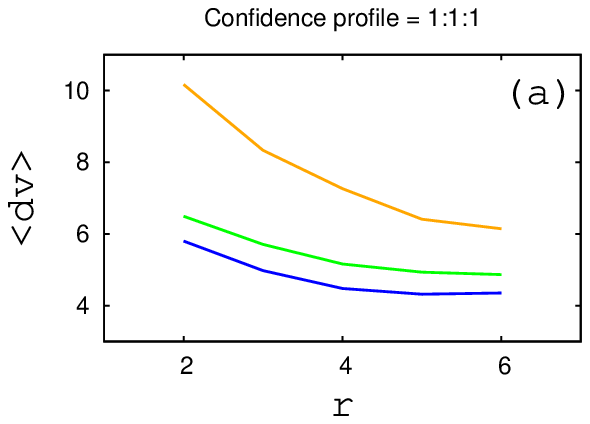}
\includegraphics[width=0.5\textwidth]{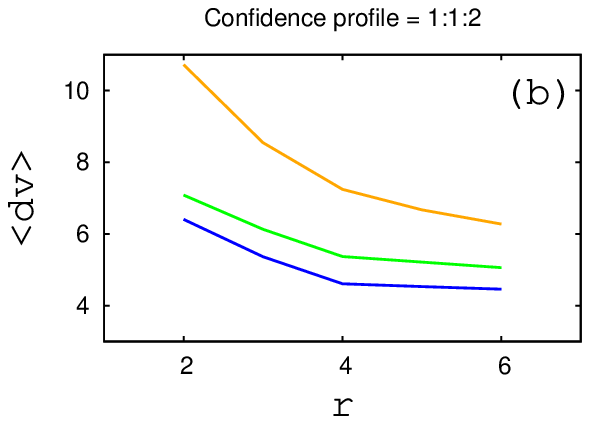}
\includegraphics[width=0.5\textwidth]{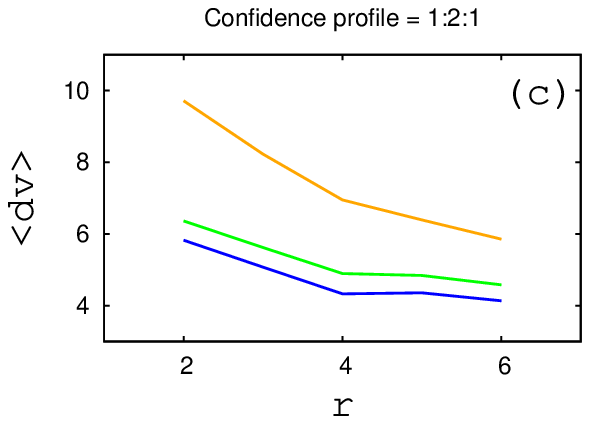}
\includegraphics[width=0.5\textwidth]{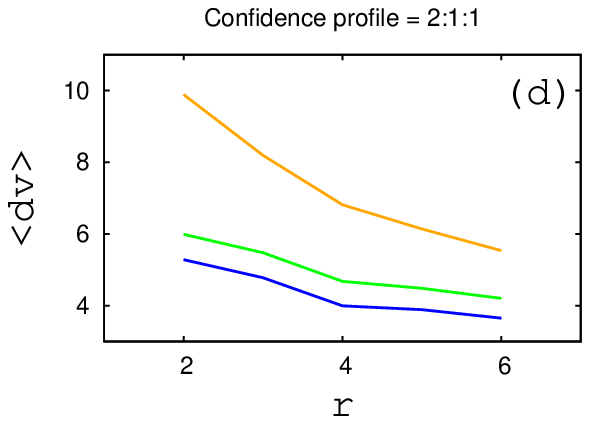}
\includegraphics[width=0.5\textwidth]{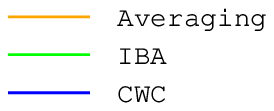}
 \vspace{-2cm}
\caption{Mean errors plotted against the number $r$ of assessors per object for the simple averaging approach (upper curves, {orange}), 
the incomplete-block-analysis method (middle curves, {green}) and the calibration-with-confidence approach (lower curves, {blue}). 
The various panels represent different confidence profiles with probabilities for high, medium and low confidences in the ratios (a) 1:1:1, (b) 1:1:2, (c) 1:2:1, (d) 2:1:1.
 }
\label{fig2}
\end{figure}

\begin{figure}[t!]
\includegraphics[width=0.5\textwidth]{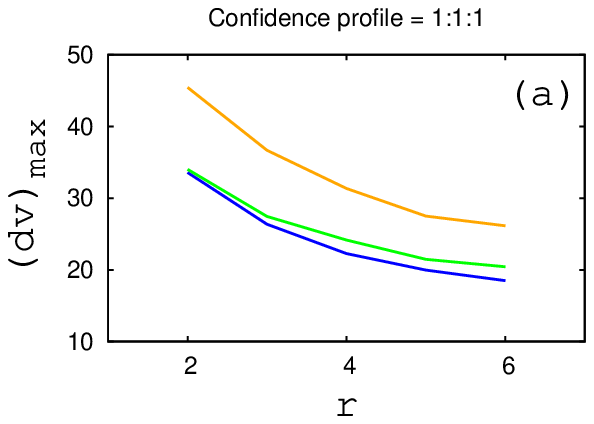}
\includegraphics[width=0.5\textwidth]{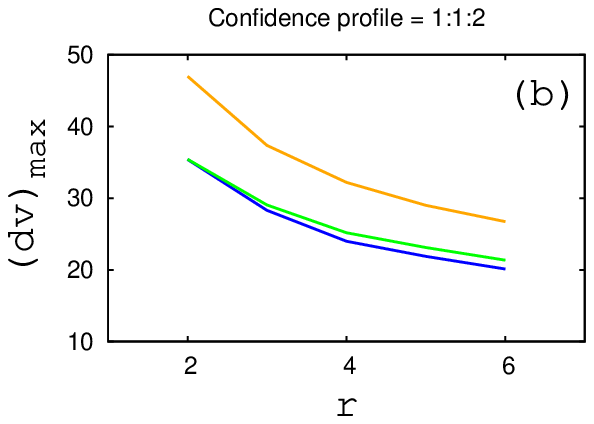}
\includegraphics[width=0.5\textwidth]{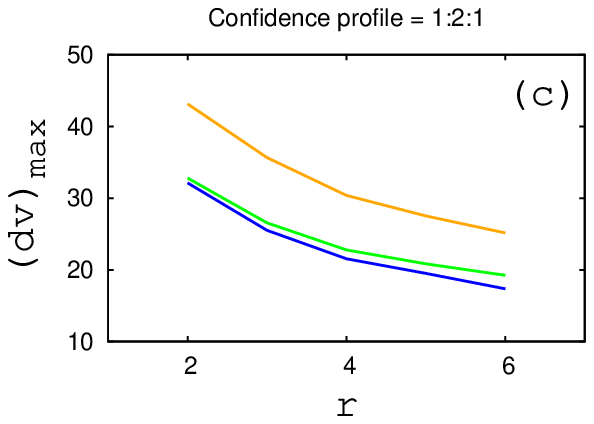}
\includegraphics[width=0.5\textwidth]{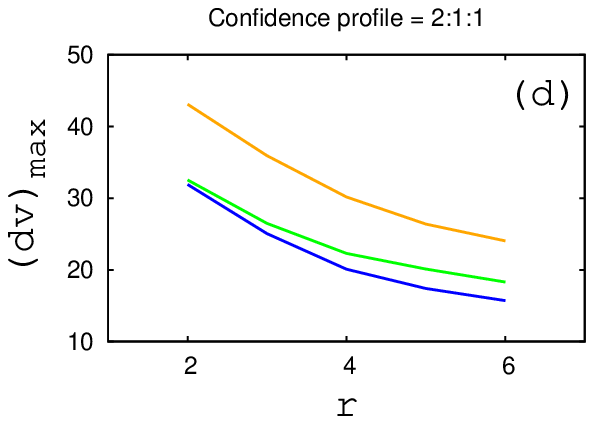}
\includegraphics[width=0.5\textwidth]{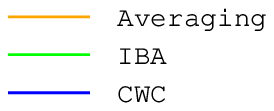}
 \vspace{-2cm}
\caption{Maximum  errors  plotted against the number $r$ of assessors per object for the simple  averaging approach (upper curves, {orange}), 
the incomplete-block-analysis method (middle curves, {green}) and the calibration-with-confidence approach (lower curves, {blue}). 
The various panels represent different confidence profiles with probabilities for high, medium and low confidences in the ratios (a) 1:1:1, (b) 1:1:2, (c) 1:2:1, (d) 2:1:1.
}
\label{fig3}
\end{figure}

\begin{figure}[t!]
\includegraphics[width=0.5\textwidth]{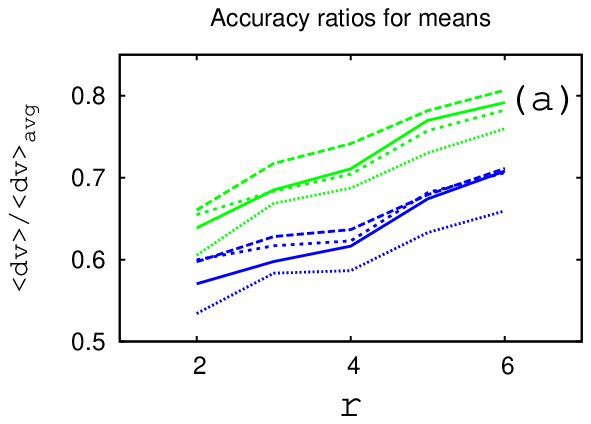}
\includegraphics[width=0.5\textwidth]{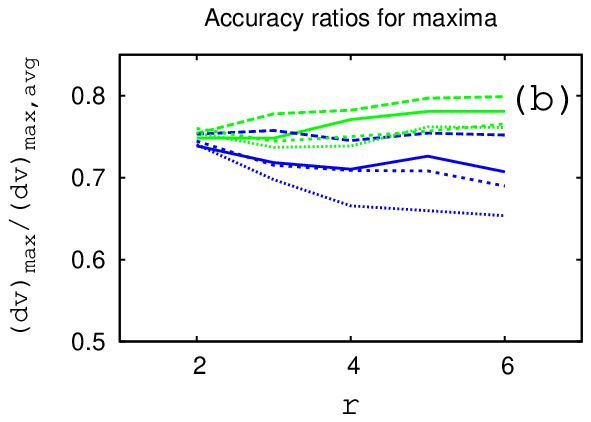}\\
\includegraphics[width=0.5\textwidth]{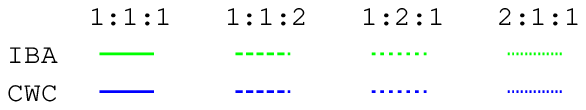}

 \vspace{-2cm}
\caption{
(a) The ratios $\langle{dv}\rangle_{\rm{{IBA}}} /  \langle{dv}\rangle_{\rm{{avg}}}$ and  $\langle{dv}\rangle_{\rm{{CWC}}}/\langle{dv}\rangle_{\rm{{avg}}}$ measure the mean improved accuracies of IBA ({green} curves)  and CWC ({blue}), respectively, over simple averaging.
Smaller ratios indicate a greater degree of improvement over SA.
(b) The analogous quantities for maximal errors are 
 $(dv)_{\rm{{max,IBA}}} /  (dv)_{\rm{{max,avg}}}$  and $(dv)_{\rm{max,CWC}}/(dv)_{\rm{max,avg}}$, respectively.
{The four line types correspond to relative probabilities 
of standard deviations of 5, 10 or 15 respectively in the ratios
1:1:1 (solid lines);  1:1:2 (long-dashed); 1:2:1 (short-dashed) and 
2:1:1 (dotted).}
 }
\label{fig4}
\end{figure}

Results for all three methods are presented in Figs.~\ref{fig2}--\ref{fig4}.
The mean and maximal  errors for the  simple averaging approach, the IBA method and the CWC approach are given in Panels (a)-(d) of Figs.~\ref{fig2} and~\ref{fig3}.
For demonstration purposes, we use  three 
confidence levels rather than a continuous distribution. This allows us to clearly control differences in confidence levels in Figs.~\ref{fig2} and~\ref{fig3} and we do so by presenting four panels labeled (a),(b),(c) and (d). 
These represent different  profiles, with the confidence for each assessment randomly allocated using probabilities for high, medium and low confidences in the ratios (a) 1:1:1, (b) 1:1:2, (c) 1:2:1, (d) 2:1:1.
We observe  that, for each method, the scores become more accurate (errors decrease) 
as the number of assessors per object $r$ increases.

From Fig.~\ref{fig2}(a)-(d), with only two assessors per object, the simple averaging method gives errors
averaging about 10 points.
Over $r=6$ assessors per object are required  to bring the mean error down to 6 {points}.
Fisher's IBA, however,  achieves this level of improvement with only 2 or 3 assessors.
The CWC method  delivers a further level of improvement of about one {point.}
One also notes that, for the calibration approaches, relatively little is gained on average by employing more than four assessors per object. 
\blue{This result can be compared with \cite{Sn} who found that five assessors per object was optimal in terms of accuracy over cost, for a procedure used by the Canadian Institutes of Health Research.}

Fig.~\ref{fig3} shows that IBA also leads to significant improvements in the maximal error values relative to those obtained through simple averaging. With two  assessors per object, maximal errors are reduced from about {45 to 30-35.} The CWC approach does not appear to significantly improve upon this. 
However, with 6 assessors per object the maximal error value of about {25} delivered by the simple averaging process is reduced to about {20} by IBA and to as low as {16} by CWC when half the assessments are done with a high degree of confidence in the scores. 

Fig.~\ref{fig4} panel~(a) gives the improvements achieved by the calibration methods as  ratios of the mean errors coming from Fisher's IBA approach to the  simple averaging approach $\langle{dv}\rangle_{\rm{IBA}} /  \langle{dv}\rangle_{\rm{avg}}$ and  of the mean errors coming from the CWC approach  to the simple averaging approach $\langle{dv}\rangle_{\rm{CWC}}/\langle{dv}\rangle_{\rm{avg}}$.
Smaller ratios mean greater accuracy on the part of the calibrated approaches.
Fig.~\ref{fig4} panel~(b) gives the analogous  accuracy ratios for the maximal errors, namely $(dv)_{\rm{{max,IBA}}} /  (dv)_{\rm{{max,avg}}}$  and $(dv)_{\rm{max,CWC}}/(dv)_{\rm{max,avg}}$. 
Fig.~\ref{fig4}(a) demonstrates that {{IBA}} delivers mean errors between about 60\% and 80\% of those coming from the simple averaging approach, the better improvements being associated with lower assessor numbers. 
This is also the most desirable configuration for realistic assessments, as it represents employment of a minimal number of assessors per object.
The CWC approach reduces errors by about a further 10 percentage points irrespective of the number of assessors.

\subsection{Case Study 2 -- Grant Proposals}
To test CWC in a realistic setting, we adapted data from a university's internal
competition for research funding, in which 43 proposals were evaluated by a panel of 11
assessors. Each proposal was graded by two assessors, who in addition each specified
a confidence-level in their grading in the form of high, medium or low.
{To respect confidentiality of the competition while making the data available, we not only anonymised the proposals and assessors but also
made sufficient changes to the data (while preserving the statistical properties) 
so that attribution would not be possible. }
The actual panel used simple averaging, but the assessors were also asked to provide confidences
so that CWC could be applied for comparison. The panel awarded grants to the top ten proposals.
Our goals were firstly to see what differences would have been made by use of IBA
or CWC, secondly to quantify the evidence for the three models from the data to
determine which was most appropriate, and thirdly to compare the posterior uncertainties they provide.

To apply CWC we translated the qualitative confidence-levels of high, medium and
low to values $c_{ao} = \lambda^2, 1,\lambda^{-2}$, respectively, with $\lambda = 1.75$. 
We chose $\lambda = 1.75$ as a reasonable guess at how the assessors used the confidence scale.
One could include a computation to infer $\lambda$ from the data, but our preference is for panel chairs to ask
assessors to provide uncertainties rather than qualitative confidence levels\blue{, as indicated in Section~\ref{sec:model},} so we did not implement the inference of $\lambda$.

\begin{figure}[t!]
\begin{center}
\includegraphics[width=0.45\textwidth]{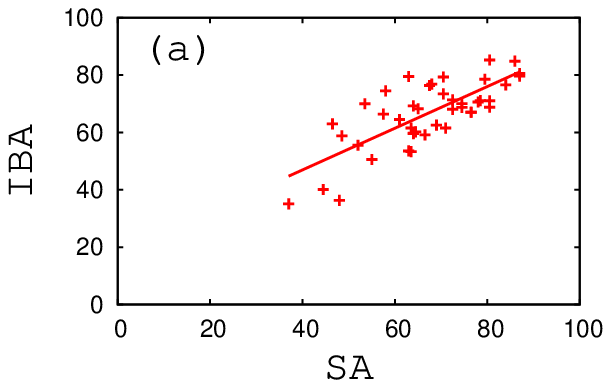}
\includegraphics[width=0.45\textwidth]{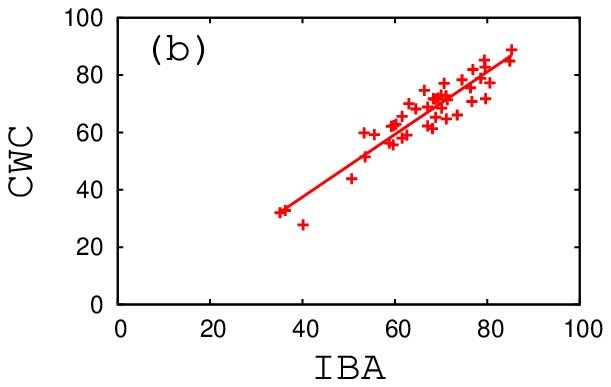}\\
\includegraphics[width=0.45\textwidth]{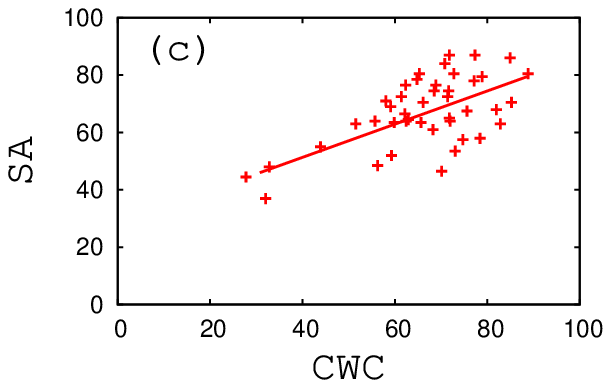}
\includegraphics[width=0.45\textwidth]{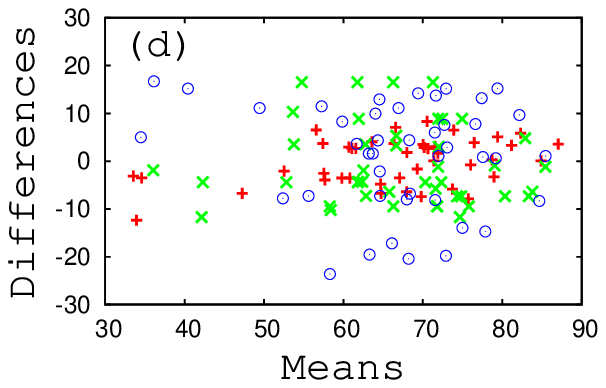}
\caption{Correlations between the results coming from the three methods applied to Case Study 2.
The three panels give the correlations between the outputs of (a) IBA and SA;
(b) CWC and IBA; (c)  SA and CWC.
The coefficients of determination are given respectively by $R^2=0.5701$; 0.8807 and 0.3772.
\red{Panel (d) is a Bland-Altman or Tukey mean-difference plot of differences between between results from pairs of approaches against their averages.
The symbols 
``$+$'' (red) compare 
CWC to IBA ($\mathcal{V}_{\rm{CWC}} - \mathcal{V}_{\rm{IBA}}$ vs $( \mathcal{V}_{\rm{IBA}} + \mathcal{V}_{\rm{CWC}})/2 $);
``$\times$'' (green) compare 
IBA to SA ($\mathcal{V}_{\rm{IBA}} - \mathcal{V}_{\rm{avg}}$ vs $( \mathcal{V}_{\rm{avg}} + \mathcal{V}_{\rm{IBA}})/2 $);
``$\circ$'' (blue) 
compare SA to CWC ($\mathcal{V}_{\rm{avg}} - \mathcal{V}_{\rm{CWC}}$ vs $( \mathcal{V}_{\rm{CWC}} + \mathcal{V}_{\rm{avg}})/2 $).}
}
\label{fig5}
\end{center}
\end{figure}

Figure~\ref{fig5} (panels a,b, and c) shows the resulting values inferred by the three methods, projected
into the planes of (SA; IBA), (IBA; CWC) and (CWC; SA). 
\red{Panel d of the same figure is a Bland-Altman or Tukey mean-difference plot
\cite{BA}.}
The correlations are not
strong, though as we would expect, the correlation of IBA with CWC is stronger
than those of either with SA. In particular, we note that the set of proposals rated 
in the top ten varies substantially with the method used (Table~\ref{ranks}). 
The reason for the differences is that IBA and CWC attribute a
significant range of biases to the assessors (Table \ref{Tbias}).

\begin{table}[b!]
\begin{center}
  \begin{tabular}{ | c || c | c | c | }
    \hline
		 {\bf{Rank}}   &  {\bf{SA $\mathcal{V}_{\rm{avg}}$}}   &  {\bf{IBA  $\mathcal{V}_{\rm{IBA}}$}}  & {\bf{CWC   $\mathcal{V}_{\rm{CWC}}$}}   \\ \hline\hline
          {\bf{1}}      &  OH	(87.0)                                     &  OA	(85.3)                              &    OA  (88.8)      \\ 
          {\bf{2}}      &  {{{\underline{OP}}}}	(87.0)         &  OC	(84.9)                              &    {\emph{OB}}	(85.2)      \\ 
          {\bf{3}}      &  OC	(86.0)                                     &  OH	(80.6)                              &    OC	(84.9)      \\ 
          {\bf{4}}      &  {{{\underline{OS}}}}	(84.0)         &  {{{\underline{OP}}}}	(79.7)  &    {\emph{OD}}	(82.8)      \\  
				  {\bf{5}}      &  OA	(80.5)                                     &  {\emph{OD}}	(79.5)                              &    {\emph{OE}}	(82.0)      \\ 
         {\bf{6}}      &  {{{\underline{OM}}}}	(80.5)         &  {\emph{OB}}	(79.4)                              &    OF	(78.9)      \\ 
          {\bf{7}}      &  {{{\underline{OZ}}}}	(80.5)         &  OF	(78.6)                              &    {\bf{{\emph{OG}}}}(78.4)      \\ 
          {\bf{8}}      &  OF	(79.5)                                     &  {\emph{OE}}	(76.9)                              &    OH	(77.3)      \\ 
          {\bf{9}}      & {{{\underline{OA}}}}$^\prime$	(78.5) &  {{{\underline{OS}}}}	(76.7)    &   {\bf{OI}}	(77.1)      \\ 
         {\bf{10}}      & OI	(78.0)                                       &  {\emph{OJ}}	(76.4)                              &    {\emph{OJ}}	(75.6)      \\ \hline
  \end{tabular}
\end{center}
\caption{The 43 grant proposals are identified as OA, OB, OC, \dots OZ, OA$^\prime$, OB$^\prime$, \dots OP$^\prime$, OQ$^\prime$. 
Here they are ranked according to their $\mathcal{V}_{\rm{avg}}$, $\mathcal{V}_{\rm{IBA}}$   and $\mathcal{V}_{\rm{CWC}}$   values, representing the outcomes of simple averaging, the IBA and CWC approaches.
Proposals identified by CWC as belonging to the top ten but missed by IBA  are highlighted in boldface.
Proposals identified by IBA or CWC as belonging to the top ten but missed by simple averaging  are highlighted in italics.
Proposals which are not in the CWC top ten are underlined.}
\label{ranks}
\end{table}

\begin{table}[b!]
\begin{center}
  \begin{tabular}{ | l || r | r | r | r | }
    \hline
   {\bf{Assessor}}     & {\bf{Mean}}    & {\bf{St. dev.}}     & {\bf{Bias (IBA)}} & {\bf{Bias (CWC)}}  \\ \hline \hline
   {\bf{AK}}           & 84.2	  & 16.6       &  14.6      &  17.7        \\ \hline
   {\bf{AJ}}           & 61.0	  & 19.2       &   8.7      &  12.6        \\ \hline
   {\bf{AI}}           & 64.6	  & 10.0       &   0.0      &   9.7       \\ \hline
   {\bf{AH}}           & 76.6	  &  9.1       &  10.0      &   9.1        \\ \hline
   {\bf{AG}}           & 71.9	  &  6.9       &   8.8      &   8.8        \\ \hline
   {\bf{AF}}           & 65.9	  &  5.6       &   5.7      &   2.0        \\ \hline
   {\bf{AE}}           & 72.3	  & 15.5       &   2.8      &   1.1        \\ \hline
   {\bf{AD}}           & 61.0	  & 21.9       & - 5.0      &   -3.6       \\ \hline
   {\bf{AC}}           & 62.3	  &  9.6       & -12.4      &  -15.6        \\ \hline
   {\bf{AB}}           & 58.3	  &  6.4       & -12.8      &  -16.6        \\ \hline
   {\bf{AA}}           & 49.1	  & 12.1       & -20.7      &  -25.2        \\ \hline
  \end{tabular}
\end{center}
\caption{Assessor statistics: 
Assessors are labeled AA, \dots AK according to increasing CWC-biases (5th column).
Here we give the mean scores they awarded, standard deviations and IBA-biases too.
The mean score awarded over all assessments was 66.9.}
\label{Tbias}
\end{table}

In the absence of ``true'' values for the proposals, how can one decide which is the
best method to use, and hence which outcome is preferred?

A first answer is to compare the ``residuals'' that the methods leave after the least squares fit. In the case of SA this means the value of (\ref{sumsquares}) obtained by taking the $v_o$ to be the averaged scores and $b_a = 0$.  For IBA, the residual is the value of (\ref{sumsquares}) at the least squares fit, taking all the $c_{ao}=1$.  For CWC, we take the value of (\ref{sumsquares}) at the least squares fit, divided by the average confidence over all assessments.  The residuals are presented in Table~\ref{resid}.  
\begin{table}[t!]
\begin{center}
  \begin{tabular}{|c |c|c|c | }
    \hline
\textbf{Method} &SA & IBA & CWC \\
\hline
\textbf{Residual} &$8602$ & $4388$ & $3156$\\
\hline
\end{tabular}
\end{center}
\caption{Residuals (scaled by mean confidence in the case of CWC).}
\vspace{1ex}
\label{resid}
\end{table}
From this point of view, we see clear improvement progressively from SA to IBA to CWC, providing an apparently compelling argument for the use of CWC.

As IBA and CWC have more free parameters (the biases) than SA, however, one should penalise them appropriately to make a correct comparison.  Also although normalising the residual for CWC by the average confidence sounds sensible, it is not clear it is the right way to compare CWC with IBA.  

A principled answer is provided by Bayesian model comparison. 
In this procedure, the evidence provided by the data in favour of each model is quantified, and the best model is the
one with the highest evidence.
\blue{T}he procedure to quantify the evidence for the three models \blue{is described} in Appendix E. It
depends on assumptions about the prior probability distribution for the parameters of
the models, but we took ``ball'' priors on the true values and on the biases (constrained by the degeneracy-breaking condition) and a truncated Jeffreys' prior on the variance of the noise.  In the notation of Appendix E, the parameters for the prior probability distributions were $\sigma_O = 22.5$, $\sigma_A = 15$, $w_{max} = 900$, $w_{min} = 1$.
As the evidences come out to be small numbers (around $10^{-168}$), we took their (natural) logarithms.
The resulting log-evidences are shown in Table \ref{bayes}.
\begin{table}[t!]
\vspace{2ex}
\begin{center}
  \begin{tabular}{|c |c|c|c | }
    \hline
\textbf{Method} &SA & IBA & CWC \\
\hline
\textbf{log-Evidence} &$-385$ & $-389$ & $-387$\\
\hline
\end{tabular}
\end{center}
\caption{Bayesian log-Evidences.}
\vspace{1ex}
\label{bayes}
\end{table}
Simple averaging wins, but these values are so close together that we can not make a strong conclusion about which method is most justified by the data.  Furthermore, adjusting the prior probability distributions and the confidence weights changes which method has the highest evidence.  
We suspect that differences between the evidences for the models would become apparent if  \blue{each proposal} had \blue{been} evaluated \blue{by} more than two  \blue{assessors}.

A third approach is to evaluate the posterior uncertainty in the values assigned to the objects for the three methods, as detailed in Appendix D, using (\ref{eq:sigma}) for IBA and CWC, and (\ref{eq:sigmaSA}) for SA.  The results are given in Table~\ref{table:uncertainties}.
\begin{table}[t!]
\vspace{2ex}
\begin{center}
  \begin{tabular}{|c |c|c|c | }
    \hline
\textbf{Method} &SA & IBA & CWC \\
\hline
\textbf{Uncertainty} &$14.1$ & $8.4$ & $8.0$\\
\hline
\end{tabular}
\end{center}
\caption{Confidence-weighted root mean-square uncertainties for the values (and biases in the cases of IBA and CWC).  For SA, the weighting is according to the number $n_o$ of assessors for object $o$.}
\label{table:uncertainties}
\end{table}
On this basis, the most precise results are given by CWC.
None of them are very precise, however.  A posterior uncertainty of 8 means that we should consider values for the objects to have a $\frac13$ chance of differing by more than 8 from the outputted values.  This means that for IBA and CWC, only the top three proposals of Table~\ref{ranks} are reasonably assured of being in the top ten.

As the object of the competition was only to choose the best 10 proposals to fund, rather than assign values to each proposal, it might have been more appropriate to design just a classifier system (with a tunable parameter to make the right number in the ``fund'' class) but our goal was to use it as a test of CWC.

The fact that three different methods with roughly equal evidence lead to drastically different allocation of the grants, and with large posterior uncertainties, highlights that better design of the panel assessment was required.
\blue{Large variability of outcome even when just using SA but with different assessment graphs was already noted by \cite{G+}.}
A moral of our analysis is that to achieve a reliable outcome, the assessment procedure needs substantial advance design.  We continue a discussion of design in Appendices C and F\blue{, but substantial treatment is deferred to a future paper}.


\subsection{Third Context -- Assessment of students}
We also tested the method 
on undergraduate examination results for a degree with a flexible options system \cite{P} and on the assessment of a multi-lecturer postgraduate module.   

In the former case, as surrogates for the confidences in the marks  we took the number of Credit Accumulation and Transfer Scheme (CATS) points for the module, which indicate the amount of time a student is expected to devote to the module (for readers used to the European credit transfer and accumulation system, 2 CATS points are equivalent to 1 ECTS point).  The amount of assessment for a module is proportional to the CATS points.  If it can be regarded as consisting of independent assessments of subcomponents, e.g.~one per CATS point, with roughly equal variances, then the variance of the total score would be proportional to the number of CATS points. 
As the score is then normalised by the CATS points, the variance becomes inversely proportional to the CATS points, making confidence directly proportional to CATS points.
The outcome \blue{of our analysis} indicated significant differences in standards for the assessment of different modules, but as most modules counted for 15 or 18 CATS, this was not a strong test of the merits of including confidences in the analysis, so we do not report on it here.  

For the postgraduate module, there were four lecturers plus module coordinator, who each assessed oral and written reports for some but not all of the students, according to availability and expertise (except the coordinator assessed them all).  Each assessor provided a score and an uncertainty for each assessment.  The results were combined using our method and the resulting value for each student was reported as the final mark.  The lecturers agreed that the outcome was fair.

\section{Discussion}

We have presented and tested a method to calibrate assessors \blue{in a panel}, taking account of differences in confidence that they express in their assessments.  
From a test on simulated data we found that Calibration with Confidence (CWC) generated closer estimates of the true values than Additive \blue{I}ncomplete \blue{B}lock \blue{A}nalysis \blue{(IBA)} or Simple \blue{A}veraging \blue{(SA)}.  A test on some real data\blue{, however,}  \blue{provided little evidence to distinguish between the methods, though they produced wildly different rankings}, suggesting that the assessment procedure for that context needed more robust design.  
Nevertheless, CWC came ahead on posterior precision.
\blue{We note that the default of assuming all assessment confidences to be equal results in IBA, which already represents a useful improvement over SA.}

\blue{One of the principal conclusions from our analysis is that to achieve reliable outcomes from the methods we tested, requires good design of the assessment graph (showing which objects are evaluated by which assessors and with what confidences).}

\blue{All three methods we compared are based on least squares fitting.  They may therefore be considered overly sensitive to outliers.  An alternative approach which is less sensitive to outliers is based on medians rather than means.  For example, Tukey's Median Polish \cite{T} is a median-based version of Fisher's IBA.  It would be good to develop a version of it that takes confidences into account too.}

\blue{Some other drawbacks of our CWC method are:
\begin{itemize}
\item it requires assessors to give reliable uncertainties; if assessors differ in their confidence estimates the method gives higher weight to those who give higher confidences.  In particular, one needs to guard against an assessor giving unwarrantedly high confidence for a particular assessment.  There is a case for calibrating confidences too.
\item bias effects may be more subtle than just an additive effect; for example an assessor may be more generous (or perhaps tougher) on topics in which they have high confidence, or they may use a shorter or longer part of the scale than other assessors.
\item some organisations insist on round-number scores; this goes against the spirit of our approach and is awkward for assessors who may rightly wish to rate an object as between two of the allowed grades.  The requirement is perhaps based on the laudable idea of not wishing to imply higher accuracy than is warranted, yet in our opinion this is better dealt with by reporting an uncertainty for each result on a continuous scale. 
\item some organisations may insist that scores can not go beyond certain limits, which is awkward for an assessor if after evaluating several objects highly they find there are some they wish to rate even higher.
\end{itemize}
}

There are a number of refinements which one could introduce to the core method\blue{, addressing some of these drawbacks}.
These include how to deal with different types of bias, different scales for confidence, different ways to remove the degeneracy in the equations, how to deal with the endpoints on a marking scale, and how to choose the assessment graph.
Some suggestions are made in the Appendices, along with mathematical treatment of the robustness of the method and of computation of the Bayesian evidence for the models.

An advantage of  \blue{our} type of calibration is that it does not produce the artificial discontinuities across field boundaries that tend to arise if the domain is partitioned into fields and evaluation in each field carried out separately.
\blue{In the UK Research Assessment Exercise 2008 for example, there is evidence that different panels had different standards \cite{KB}.  Although RAE2008 stated that cross-panel comparisons are not justified, some universities have used such comparisons to help decide on how much to resource different departments.  Our approach would take advantage of cross-panel referrals (which was part of RAE2008 for work in the overlaps between panels) to infer relative standards and hence to normalise the outcomes.}

We  suggest that a method such as this, which takes into account declared confidences in each assessment, is well suited to a multitude of situations in which a number of objects is assessed by a panel. 
\blue{We acknowledge, however, that this approach requires an investment in training assessors to estimate their uncertainties and in constructing a sufficiently strongly connected assessment graph.  Different panels will deal with the trade-off between investment of effort and accuracy of results in different ways.}

\vspace{0.5cm}
\section*{Acknowledgements}
We are grateful to the Mathematics Department, University of Warwick, for providing us with examination data to perform an early test of the method, to the Applied Mathematics Research Centre, Coventry University for funding to make a professional implementation of the method and to Marcus Ong and Daniel Sprague of Spectra Analytics for producing it.
We also thank John Winn for pointing us to the SIGKDD'09 method, and David MacKay for pointing us to the NIPS method\blue{ and teaching RM Bayesian model comparison back in 1990}.  \blue{We are grateful to the reviewers of this and previous versions for many useful comments.}

\vspace{0.5cm}
{\section*{Data Accessibility}}

Software implementing the method is free to download from the website\\
{\tt{http://calibratewithconfidence.co.uk}}. 
Software and data for  the two case studies are available from  
{\tt{https://github.com/ralphkenna/CWC.git}}.

\vspace{0.5cm}
\section*{Author Contributions}
RM conceived and developed the theory.
SP tested it using an early case study.
RL performed case study 1.
RK performed case study 2.
RM,  SP, RK and  RL discussed and interpreted the results and wrote the paper.

{
\vspace{0.5cm}
\section*{Funding statement}

The work of RM was supported by the ESRC under the Network on Integrated Behavioural Science (ES/K002201/1) and the Centre for Evaluation of Complexity in the Nexus (ES/N012550/1).
RK was supported by the EU Marie Curie IRSES Network PIRSES-GA-2013-612707
DIONICOS - Dynamics of and in Complex Systems
funded by the European Commission within the FP7-PEOPLE-2013-IRSES Programme (2014-2018).
}

{
\vspace{0.5cm}
\section*{Competing Interests}

We have no competing interests.
}


\vspace{0.5cm}
\section*{Appendix A: Scale for confidences}

We motivated the model by proposing that the noise terms be of the form $\sigma_{ao} \eta_{ao}$ with the $\eta_{ao}$ independent zero-mean random variables with unit variance, so that the $\sigma_{ao}$ are standard deviations.  Nevertheless, multiplying all the confidences by the same number does not change the results of the least squares fit, nor our quantifications of robustness (Appendices C and D).  Thus the $\eta_{ao}$ can be taken to have any variance $w$, as long as it is the same for all assessments.
It is only ratios of confidences that have significance.

The fitting procedure can be extended to infer a best fit value for $w$.  Even if the assessors provide confidences based on assuming $w=1$, the best fit for $w$ is not 1 in general.  Assuming independent Gaussian errors, the maximum likelihood value for $w$ comes out to be 
$$\bar{w} = R/N,$$ 
where 
\begin{equation}
R= \sum_{ao} c_{ao} (s_{ao}-\bar{v}_o-\bar{b}_a)^2
\label{eq:R}
\end{equation}
is the residual from the least squares fit ($\bar{v},\bar{b})$ for $(v,b)$ and $N$ is the total number of assessments.
The posterior distribution for $w$, given a prior distribution, is obtained in Appendix D.

\section*{Appendix B: Degeneracy-breaking conditions}

We can remove the degeneracy in the equations {(\ref{eq:system1}) and (\ref{eq:system1b})} in different manners  \blue{from} equation~(\ref{bias}) used here.   
Indeed, use of (\ref{bias}) can lead to an average shift from the scores to the true values.  This does not matter if only a ranking is required, but if the actual values are important\blue{ (e.g.~for degree classification)}, then a better choice of degeneracy-breaking condition is needed.

A preferable confidence-weighted degeneracy-breaking condition is 
\begin{equation}
\sum_a C'_a b_a=0, 
\label{eq:avbias}
\end{equation}
which from (\ref{eq:system1b}) automatically implies $\sum_o C_o v_o = \sum_{ao} c_{ao} s_{ao}$, thus avoiding the possibility of such systematic shifts.  

From a theoretical perspective, however, the best choice of degeneracy-breaking condition is to choose a reference value $v_{\rm ref}$ (think of a notional desired mean) and require
\begin{equation}
\sum_{ao} c_{ao}(v_o-b_a) = C v_{\rm ref} ,
\label{tightest}
\end{equation}
where
\begin{equation}
C = \sum_{ao} c_{ao} .
\end{equation}
Using the notation in (\ref{eq:vob}) and (\ref{eq:voc}) this can equivalently be written as
\begin{equation}
\sum_o C_o v_o - \sum_a C'_a b_a = C v_{\rm ref} .
\label{eq:tightest}
\end{equation}
To reduce the possible average shift from confidence-weighted average scores to true values, the reference value $v_{\rm ref}$ should be chosen near the confidence-weighted average score 
\begin{equation}
\bar{s} = \sum_{ao} c_{ao} s_{ao}/C. 
\end{equation}
Choosing $v_{\rm{ref}}$ exactly equal to $\bar{s}$ gives (\ref{eq:avbias}), which makes the confidence-weighted average bias come out to 0 and the confidence-weighted average value come out to $\bar{s}$.  
We will show in Appendix C, however, that the results are a factor $\sqrt{2}$ more robust to changes in the scores if $v_{\rm ref}$ is chosen to be fixed rather than dependent on the scores.

\blue{For any affine choice of degeneracy-breaking condition on the biases, $\sum_a \beta_a b_a = \gamma$, the reduced system (\ref{eq:reduced}) can be solved either by replacing one of the equations by the degeneracy-breaking condition as in Section~\ref{sec:solution}, or by appending an additional unknown $s$, adding $\beta_a s$ to the lefthand side of each equation (\ref{eq:reduced}), and appending the degeneracy-breaking equation as an additional equation.  The latter option has the advantage of preserving the symmetry of the matrix representing the system of equations and hence twice as efficient algorithms to solve them (symmetric indefinite factorisation).  The additional unknown $s$ comes out to be $0$ because of the relation $\alpha(V,B)=0$ mentioned after (\ref{eq:alpha}).}

\section*{Appendix C: Robustness to changes in the scores}

Here we present our approach to the quantification of the robustness of our method to small changes in the scores, using norms that take into account the confidences.



For $s = (s_{ao})_{(a,o) \in E}$, define the operator $K$ by
\begin{equation}
K s = \left[ \begin{array}{c} V \\ B \end{array} \right] ,
\end{equation}
as a shorthand for the definitions in equations~(\ref{Referee1a}) and (\ref{Referee1b}),  so that the equations (\ref{eq:system1}, \ref{eq:system1b}) can be written as 
\begin{equation}
L \left[ \begin{array}{c} v \\ b \end{array} \right] = K s .
\end{equation}
Thus, if a change $\Delta s$ is made to the scores, we obtain changes $\Delta v$, $\Delta b$ of magnitude bounded by
\begin{equation}
\left\| \begin{array}{c} \Delta v \\ \Delta b \end{array} \right\| \le \|L^{-1}K\| \|\Delta s \| ,
\label{bound}
\end{equation}
where $L^{-1}$ is defined by restricting the domain of $L$ to (\ref{bias}) and its range to $\alpha(V,B)=0$, and appropriate norms are chosen.  In this appendix, we propose that appropriate choices of norms are
\begin{equation}
\| \Delta s \|_{\rm{scores}} = \sqrt{\sum_{ao} c_{ao}\ \Delta s_{ao}^2 },
\label{eq:scoresnorm}
\end{equation}
\begin{equation}
\| (\Delta v, \Delta b) \|_{\rm{results}} = \sqrt{\sum_{ao} c_{ao} (\Delta v_o^2 + \Delta b_a ^2)} 
= \sqrt{\sum_o C_o\ \Delta v_o^2 + \sum_a C'_a\ \Delta b_a^2} ,
 \label{eq:resultsnorm}
\end{equation}
and the associated operator norm from scores to results for $\|L^{-1}K\|$.
With the  confidence-weighted degeneracy-breaking condition $\sum C'_a b_a = 0$ (\ref{eq:avbias}) instead of (\ref{bias})
we obtain
\begin{equation}
\| L^{-1} K \| \le \frac{\sqrt{2}}{\sqrt{\mu_2}},
\end{equation}
where $\mu_2$ is the second smallest eigenvalue of a certain matrix $M$ formed from the confidences (see (\ref{eq:M})).
In particular, this gives
\begin{equation}
| \delta v_o | \le \frac{1}{\sqrt{C_o}} \frac{\sqrt{2}}{\sqrt{\mu_2}} \sqrt{\sum_{ao} c_{ao}\ \delta s_{ao}^2} .
\end{equation}
The factor of $\sqrt{2}$ can be removed if one switches to an ideal degeneracy-breaking condition as in (\ref{tightest}) of Appendix B.

As a consequence, to maximise the robustness of the results, the task for the designer of $E$ is to make none of the $C_o$ much smaller than the others and to make $\mu_2$ significantly larger than 0.  The former is evident (no object should receive significantly less assessment or less expert assessment than the others).  The latter is the mathematical expression of how well connected is the graph $\Gamma$ (equivalently $\Gamma_A$).  To design the graph $\Gamma$ requires a guess of the confidence levels that assessors are likely to give to their assessments (based on knowing their areas of expertise and their thoroughness or otherwise) and a compromise between assigning an object to only the most expert assessors for that object and the need to achieve a chain of comparisons between any pair of assessors.


We now go into detail, derive the above bounds and describe some computational shortcuts.

One can measure the size of a change $\Delta s_{ao}$ to a score $s_{ao}$ by comparing it to the declared uncertainty $\sigma_{ao}$.  Thus we take the size of $\Delta s_{ao}$ to be $\sqrt{c_{ao}}\ |\Delta s_{ao}|$.
We propose to measure the size of an array $\Delta s$ of changes $\Delta s_{ao}$ to the scores by the square root of the sum of squares of the sizes of the changes to each score, as in (\ref{eq:scoresnorm}).
Supremum or sum-norms could also be considered but we will stick to this choice here.

It is also reasonable to measure the size of a change $\Delta v_o$ to a true value $v_o$ by comparing it to the uncertainty implied by the sum of confidences in the scores for object $o$.  Thus the size of $\Delta v_o$ is defined to be
$\sqrt{C_o}\ |\Delta v_o | $, where $C_o$ is the total confidence in the assessment of object $o$.
Similarly, we measure the size of a change $\Delta b_a$ in bias $b_a$ by 
$\sqrt{C^\prime_a}\ | \Delta b_a |$ where $C^\prime_a$ is the total confidence expressed by a given assessor.
Finally, we measure the size of a change $(\Delta v, \Delta b)$ to the vector of values and biases by the square root of sum of squares of the individual sizes, as in (\ref{eq:resultsnorm}).
 
The size of the operator $L^{-1}K$ is measured by the operator norm from scores to results, i.e.
\begin{equation}
\|L^{-1}K\| = \sup_{\Delta s \ne 0} \frac{\|L^{-1}K \Delta s\|_{\rm results}}{\|\Delta s\|_{\rm scores}}.
\end{equation}
The operator $L^{-1}K$ is equivalent to orthogonal projection with respect to the norm (\ref{eq:scoresnorm}) from the scores to the subspace $\Sigma$ of the form $s_{ao} = v_o + b_a$ with a degeneracy-breaking condition to eliminate the ambiguity in direction of the vector $v_o=1, b_a = -1$.  
 
The tightest bounds in (\ref{bound}) are obtained by choosing the degeneracy-breaking condition to correspond to a plane perpendicular to this vector with respect to the inner product corresponding to equation~(\ref{eq:resultsnorm}).  Thus we choose degeneracy-breaking condition (\ref{tightest}).

 \vskip 1ex
\noindent{\bf Theorem}: For a connected graph $\Gamma$ and with the degeneracy-breaking condition (\ref{tightest}), the size of the change $(\Delta v, \Delta b)$ resulting from a given array of changes $\Delta s$ in scores is bounded by
 \begin{equation}
 \| (\Delta v, \Delta b) \|_{\rm{results}} \le \frac{1}{\sqrt{\mu_2}} \| \Delta s \|_{\rm{scores}},
 \end{equation}
 where $\mu_2$ is the second smallest eigenvalue of the matrix
\begin{equation}
M = \left[ \begin{array}{cc}
 I_{N_O} & D \\
 D^T & I_{N_A}
 \end{array}\right],
 \label{eq:M}
 \end{equation}
\begin{equation}
D^T_{ao} =  {c_{ao}}/{\sqrt{C_oC^\prime_a}},
\end{equation} 
$N_A$, $N_O$ are the numbers of assessors and objects respectively, and for $k \in \mathbb{N}$\blue{,} $I_k$ is the identity matrix of rank $k$.
 
 \vskip 1ex
\noindent{\bf Proof}: Firstly, the orthogonal projection in metric (\ref{eq:scoresnorm}) from $s$ to the subspace $\Sigma$ never increases length.  
Secondly, if $\Delta s_{ao} = \Delta v_o+\Delta b_a$ with $\sum_{ao} c_{ao} (\Delta v_o-\Delta b_a)=0$ then 
\begin{equation}
\| \Delta s\|_{\rm{scores}}^2 = \sum_{ao} c_{ao} (\Delta v_o+\Delta b_a)^2  = g^TMg ,
\label{eq:gTMg}
\end{equation}
where $g$ is the vector with components 
\begin{eqnarray}
g_o &=& \tilde{v}_o := \sqrt{C_o}\ \Delta v_o, \\
g_a &=& \tilde{b}_a := \sqrt{C'_a} \ \Delta b_a.
\end{eqnarray}
Then, because we restricted to the orthogonal subspace to the null vector in results-norm and $M$ is non-negative and symmetric,
$$ g^T Mg \ge \mu_2 \sum_i g_i^2 = \mu_2 \|(\Delta v,\Delta b)\|^2_{\rm{results}},$$
where index $i$ ranges over all objects and assessors.  Positivity of $\mu_2$ holds as soon as the graph $\Gamma$ is connected, because $M$ is a transformation of the weighted graph-Laplacian to scaled variables \cite{Ch}, so dividing by $\mu_2$ and taking the square root yields the result. $\Box$

\vskip 1ex

The computation of the eigenvalue $\mu_2$ of $M$ can be reduced from dimension $N_A+N_O$ to dimension $N_A$ by

\vskip 1ex
\noindent{\bf Proposition}: If $N_A \ge 2$, the second smallest eigenvalue $\mu_2$ of $M$ is related to the second largest eigenvalue $\lambda_2$ of $D^T D$ by
\begin{equation}
\mu_2 = 1 - \sqrt{\lambda_2} .
\end{equation}
If $N_A = 1$ and $N_O \ge 2$ then $\mu_2 = 1$.  If both are 1 then $\mu_2 = 2$.

\vskip 1ex
\noindent{\bf Proof}: The equations for an eigenvalue-eigenvector pair $\mu, (\tilde{v},\tilde{b})$ of $M$ are
\begin{eqnarray}
\tilde{v}+D\tilde{b} &=& \mu  \tilde{v} \label{eq:evec} \\
D^T \tilde{v} + \tilde{b} &=& \mu \tilde{b}.
\end{eqnarray}
Applying $D^T$ to the first equation, multiplying the second by $(1-\mu)$, and then substituting for $(1-\mu)D^T \tilde{v}$ in the second yields
\begin{equation}
D^T D \tilde{b} = (1-\mu)^2 \tilde{b}.
\end{equation}
Thus either $\tilde{b}=0$ or $(1-\mu)^2$ is an eigenvalue $\lambda$ of $D^T D$.  In the first case,  equation (\ref{eq:evec}) implies $\mu =1$, so if $\mu \ne 1$ then $(1-\mu)^2$ is an eigenvalue of $D^TD$. 

Conversely, if $(\lambda, \tilde{b})$ is an eigenvalue-eigenvector pair for $D^TD$ with $\lambda \ne 0$ then $\lambda > 0$ because $D^T D$ is non-negative, so put $\tilde{v} = \pm D\tilde{b}/\sqrt{\lambda}$ to see that $(\tilde{v},\tilde{b})$ is an eigenvector of $M$ with eigenvalue $\mu = 1\pm\sqrt{\lambda}$.  If $\lambda=0$ and $D\tilde{b} = 0$ then $\mu=1$ is an eigenvalue of $M$ with eigenvector $(\tilde{v},\tilde{b})$ for any $\tilde{v}$ with $D^T\tilde{v}=0$, e.g.~$\tilde{v}=0$.

Thus there is a two-to-one correspondence between eigenvalues $\mu$ of $M$ not equal to 1 and positive eigenvalues $\lambda$ of $D^T D$ (counting multiplicity): 
$\mu = 1\pm\sqrt{\lambda}$.
Any remaining eigenvalues are 1 for $M$ and 0 for $D^TD$.
The degeneracy gives an eigenvector $\tilde{v}_o = \sqrt{C_o}, \tilde{b}_a = -\sqrt{C'_a}$ of $M$ with eigenvalue 0 and it corresponds to an eigenvalue 1 of $D^TD$.  All other eigenvalues of $M$ are non-negative because $M$ is.  All other eigenvalues of $D^TD$ are less than or equal to 1 by the Cauchy-Schwarz inequality.  So if the second largest eigenvalue $\lambda_2$ of $D^TD$ (counting multiplicity) is positive then the second smallest eigenvalue $\mu_2$ of $M$ (counting multiplicity) is $1-\sqrt{\lambda_2}$.  If $\lambda_2 = 0$ then $\mu_2=1$ because existence of $\lambda_2$ implies $N_A\ge 2$ so $M$ has dimension at least 3 and we have only two simple eigenvalues $\mu = 0$ and $2$ from the simple eigenvalue 1 of $D^TD$, so $M$ must have another one but
any other value than 1 would give a positive $\lambda_2$; so the same formula holds.  If there is no second eigenvalue of $D^TD$ (because $N_A=1$) then if $N_O\ge 2$ the second largest eigenvalue of $M$ must be 1 by the same argument.  If both $N_A$ and $N_O$ are 1 then the second largest eigenvalue of $M$ is the other one associated with the eigenvalue 1 of $D^TD$, namely 2.
$\Box$

\vskip 1ex

Note that 
$$(D^TD)_{aa'} = C_{aa'}/\sqrt{C^\prime_a C^\prime_{a'}}$$ 
is a similarity transformation of (\ref{eq:Aweights}).  As examples of second eigenvalues, putting unit confidences on the graphs in the left column of Figure~\ref{fig1:graphs} we calculate $\lambda_2 = 1/3, 2/3, 1$ for cases (a),(b),(c) in the right column, giving $\mu_2 = 1-\sqrt{1/3}, 1-\sqrt{2/3}, 0$, respectively.

Finally, a user may prefer to use the degeneracy-breaking condition (\ref{eq:avbias})
rather than (\ref{tightest}), perhaps out of uncertainty about what value of $v_{\rm{ref}}$ to use.  Or a user may be happy to use (\ref{tightest}) with $v_{\rm{ref}}$ equal to the confidence-weighted average score, but wants $v_{\rm{ref}}$ to follow this average score if changes are made to the scores.  That comes out equivalent to using (\ref{eq:avbias}).  So we extend our discussion of robustness to treat this case.  We find it makes the bounds increase by a factor of only $\sqrt{2}$.

\vskip 1ex
\noindent{\bf Proposition}: For $\Gamma$ connected and using degeneracy-breaking condition (\ref{eq:avbias}), the size of $(\Delta v, \Delta b)$ resulting from changes $\Delta s$ to the scores is at most $\frac{\sqrt{2}}{\sqrt{\mu_2}} \|\Delta s\|_{\rm{scores}} $.

\vskip 1ex

\noindent{\bf Proof}: If the degeneracy-breaking condition (\ref{tightest}) gives a change $(\Delta v, \Delta b)$ for a change $\Delta s$ to the scores, then switching to degeneracy-breaking condition (\ref{eq:avbias}) just adds an amount $k$ of the null vector $\bf{n}=(\bf{1},\bf{-1})$ to achieve $\sum_a C'_a (\Delta b_a - k) = 0$, i.e.
\begin{equation}
k = \frac{\sum_a C'_a \Delta b_a}{C} .
\end{equation}
In the results metric, the null vector has length $\sqrt{\sum_o C_o + \sum_a C'_a} = \sqrt{2C}$.  Thus the correction has length $|k|\sqrt{2C} = \sqrt{\frac{2}{C}}\ |\sum C'_a \Delta b_a|$.
Using the condition (\ref{tightest}) we can write 
$\sum_a C'_a \Delta b_a = \frac12(\sum_a C'_a \Delta b_a + \sum_o C_o \Delta v_o),$
which one can recognise as one half of the inner product of $(\bf{1},\bf{1})$ with $(\Delta v, \Delta b)$ in results-norm, so it is bounded by $\sqrt{C/2}\ \|(\Delta v, \Delta b)\|$.
Thus the length of the correction vector is at most that of $(\Delta v, \Delta b)$.  The correction is perpendicular to $(\Delta v, \Delta b)$, thus the vector sum has length at most $\sqrt{2}\ \|(\Delta v, \Delta b)\|$.
$\Box$

\vskip 1ex

One may also ask about robustness with respect to changes in the confidences $c_{ao}$.  If an assessor declares extra high confidence for an evaluation, for example, that can significantly skew the resulting $v$ and $b$.  
The analysis is more subtle, however, because of how the $c_{ao}$ appear in the equations and we do not treat it here.

\section*{Appendix D: Posterior probability distribution}

Another point of view on robustness is the Bayesian one.  From a prior probability on $(v,b)$ and a model for the $\eta_{ao}$, one can infer a posterior probability for $(v,b)$, whose inverse width tells one how robust is the inference. 

In the case of flat prior on $(v,b)$, prescribed $w$,
Gaussian noise, and an affine degeneracy-breaking condition, the posterior is Gaussian with mean at the value solving equations~(\ref{eq:system1}), (\ref{eq:system1b}) and the degeneracy-breaking condition, and with covariance matrix related to $L^{-1}$.  Specifically, the posterior probability density for $(v,b)$ is proportional to
$$\exp -\frac{S}{2w}, $$
constrained to the degeneracy-breaking hyperplane, where
$$ S = \sum_{ao} c_{ao}(s_{ao}-v_o-b_a)^2.$$
Using (\ref{eq:gTMg}) and (\ref{eq:R}), this can be written as
$$\exp -\frac{1}{2w} (g^T M g + R),$$
with $(\Delta v, \Delta b)$ being the deviations of $(v,b)$ from the least squares fit.  Thus the covariance matrix in these scaled variables is $w M^{-1}$\blue{, where for degeneracy-breaking condition $\gamma^T g = K$, $M^{-1}$ is interpreted as the limit as $t \to \infty$ of $(M+t\gamma \gamma^T)^{-1}$}.  Using the degeneracy-breaking condition (\ref{tightest}) or equivalently (\ref{eq:tightest})\blue{ for which $\gamma$ is in the null direction of $M$ and diagonalising the matrix}, we obtain widths $\sqrt{w/\mu_j}$ for the posterior on $g$ in the eigendirections of $M$, where $\mu_j$ are the positive eigenvalues of $M$.  Thus the robustness of the inference is again determined by $\mu_2$, but scaled by $\sqrt{w}$.

A slightly more sophisticated approach is to consider $w$ to be unknown also.  Given a prior density $\rho$ for $w$ (which could be peaked around 1 if the assessors are assigning confidences via  uncertainties, but following Jeffreys would be better chosen to be $1/w$ if there is no information about the scale for the confidences), the posterior density for $(w,v,b)$ is proportional to
$$\rho(w) w^{-N/2} \exp - \frac{S}{2w} ,$$
where again $N$ is the number of assessments.
The maximum of the posterior probability density is determined by the least squares fit for $(v,b)$ (which is independent of $w$) and the following equation for $w$:
$$\frac{\rho'(w)}{\rho(w)} - \frac{N}{2w} + \frac{S}{2w^2} = 0.$$
For $N$ large, the peak of the posterior has $w$ near the previously determined maximum likelihood value $\bar{w} = R/N$.   \blue{For example}, taking Jeffreys' prior, the peak is at $w = R/(N-2)$.
Integrating over $w$ (with Jeffreys' prior) \blue{on}e find\blue{s} the marginal posterior for $(v,b)$ to be proportional to 
$$(g^T Mg + R)^{-N/2}.$$ 
Incorporating an affine degeneracy-breaking condition, this is a $(N_O+N_A-1)$-variate Student distribution with $\nu = N-N_O-N_A+1$ degrees of freedom.  Its covariance matrix is $w^* M^{-1}$ with 
$$w^* = \frac{R}{\nu-2}$$ 
and $M^{-1}$ interpreted by imposing the chosen degeneracy-breaking condition \blue{as above}.  

So for the degeneracy-breaking condition (\ref{tightest}), the robustness of the inference is given by widths $\sqrt{w^*/\mu_j}$ for $j \ge 2$, in the eigendirections of $M$ on $g$.  In particular, the confidence-weighted root mean square uncertainty $\sigma$ for the components of the vector $(v,b)$ is
\begin{equation}
\sigma = \sqrt{\frac{w^*}{2C} \sum_{j\ge 2}\frac{1}{\mu_j}} = \sqrt{\frac{R\ \Tr M^{-1}}{2(\nu-2)C}} ,
\label{eq:sigma}
\end{equation}
where $\Tr$ denotes the trace and, again, $M^{-1}$ is interpreted by restricting to the degeneracy-breaking plane.
Marginal posteriors for each $v_o$ and $b_a$ can be extracted, but it must be understood that in general they are significantly correlated. 
\blue{One way to do this in the case of degeneracy-breaking condition (\ref{tightest}) is to find the orthogonal matrix $O$ to diagonalise $M$ as $O^T D O$ with $D = diag(\mu_j)$, and then the posterior variance of $g_i$ is $w^* \sum_{j>1} O_{ji}^2/\mu_j$, but there may be ways to evaluate it without diagonalising $M$.}

For the case of simple averaging, the root mean-square posterior uncertainty in the values, weighted by the numbers $n_o$ of assessors for object $o$, is
\begin{equation}
\sigma = \sqrt{\frac{R}{N-N_O},}
\label{eq:sigmaSA}
\end{equation}
where $R$ is defined in (\ref{eq:Rsa}) of Appendix E.
This can be derived in an analogous fashion to (\ref{eq:sigma}) via a Student distribution again, but with $M = I_{N_O}$.

\section*{Appendix E: Model comparison}
Here we describe the method used in Case Study 2 to compare the three models.

Bayesian model comparison is based on computing how much evidence there is for each proposed model, e.g.~Ch.28 of \cite{Mac}.
The evidence for a model $M$ given data $D$ is $P(D|M)$.  Given strength of belief $P(M)$ in model $M$ prior to the data (relative to other models), one can multiply it by the evidence to obtain the posterior strength of belief in model $M$.
It is convenient to replace multiplication by addition, thus we define the log-evidence $$LE(M|D) = \log P(D|M).$$
If the model $M$ has free parameters $\mu$ then
$$ P(D|M) = \int P(D|M,\mu) P_M(\mu)\ d\mu, $$
where $P_M(\mu)$ is a prior probability density on $\mu$.

Let there be $N_O$ objects, $N_A$ assessors, let $s_{ao}$ be the score returned by assessor $a$ for object $o$, $c_{ao}$ the confidence in this score in the case of calibration with confidence, $s$ be the collection of scores and $N$ be their number.

First we compute the evidence for simple averaging (SA).  Then we treat calibrate with confidence (CWC) and lastly incomplete block analysis (IBA) because it is a special case of CWC.

\subsection*{-- Simple Averaging}

For simple averaging (SA), the model is that $$s_{ao} = v_o + \eps_{ao}$$ 
for some unknown vector $v$ of ``true'' values $v_o$, with $\eps_{ao}$ iid normal $N(0,w)$ for some unknown variance $w$.  Then the probability density for the scores $s$ is
$$ P(s|v,w) = \prod \frac{e^{-(s_{ao}-v_o)^2/2w}}{\sqrt{2\pi w}}
= (2\pi w)^{-N/2} e^{-\frac{1}{2w}\sum (s_{ao}-v_o)^2} ,$$
with the product and sum being over the assessments that were carried out.

To work out the evidence for SA the model must include a prior probability density for $v$ and $w$.  The simplest proposal would be
$\Delta^{-N_O} L^{-1} w^{-1}$
on $v_o \in [v_{min},v_{max}]$, $w \in [w_{min}, w_{max}]$, where $\Delta = v_{max}-v_{min}$ and $L = \log(w_{max}/w_{min})$.  This is the product of a ``box'' prior on $v$ and Jeffreys' prior on $w$ (truncated to an interval and normalised).
For comparison with the other models, however, it is easier to replace the box prior on $v$ by a ``ball'' prior, giving
$$P_{SA}(v,w) = \frac{1}{Z_OLw}$$
on 
\begin{equation}
\sum_o (v_o-v_{ref})^2 \le N_O \sigma_O^2,
\label{eq:ballO}
\end{equation}
for some anticipated average score $v_{ref}$ and upper estimate of the width $\sigma_O$ of the distribution of values $v_o$.
The normalisation is
$$Z_O =  \frac{(\pi N_O\sigma_O^2)^{N_O/2}}{\Gamma(N_O/2+1)},$$
where $\Gamma$ is the Gamma function.
For $w_{min}$ is it reasonable to choose $u^2$ where $u$ is the smallest change any assessor could contemplate.  For $w_{max}$ it is reasonable to choose $\sigma_O^2$.

For each object $o$,
$$\sum_a (s_{ao}-v_o)^2 = n_o (v_o-\bar{s}_o)^2 + R_o,$$
where $n_o$ is the number of assessors for object $o$, $\bar{s}_o$ is the mean of their scores, and the residual 
\begin{equation}
R_o =\sum_a (s_{ao}-\bar{s}_o)^2.
\label{eq:Ro}
\end{equation}
Thus 
$$P(s|v,w)P_{SA}(v,w) = \frac{1}{Z_OLw}(2\pi w)^{-N/2} e^{-\frac{1}{2w} \sum_o n_o (v_o-\bar{s}_o)^2} e^{-R/2w},$$
where
\begin{equation}
R = \sum_o R_o.
\label{eq:Rsa}
\end{equation}
To integrate this over $v$ and $w$, we assume the bulk of the probability distribution lies in the product of the ball and the interval, and so approximate by extending the range of integration to $\R^{N_O}\times (0,\infty)$.
Integrating the exponential over $v_o$ produces a factor
$$\sqrt{\frac{2\pi w}{n_o}}.$$
Thus, integrating over all components of $v$ yields
$$\frac{1}{Z_OLw} (2\pi w)^{-(N-N_O)/2}e^{-R/2w} \prod n_o^{-\frac12}.$$
Integrating this over $w$,
we obtain the evidence
$$P(SA|s) = \frac{1}{Z_OL}(\pi R)^{-(N-N_O)/2}\ \Gamma\left(\frac{N-N_O}{2}\right)\prod n_o^{-\frac12}$$
and the log-evidence
$$LE(SA|s) = \log\Gamma\left(\frac{N-N_O}{2}\right) - \frac{N-N_O}{2}\log \pi R - \frac12 \sum_o \log n_o - \log Z_O - \log L.$$

\subsection*{-- Calibration with Confidence}

For Calibrate with Confidence (CWC), the model is
$$s_{ao} = v_o + b_ a + \sigma_{ao} \eta_{ao},$$
for some unknown vectors $v$ of true values $v_o$, and $b$ of assessor biases $b_a$, with $\eta_{ao}$ iid normal $N(0,w)$ for some unknown variance $w$.  The uncertainties $\sigma_{ao}$ correspond to confidences $c_{ao}$ by $\sigma_{ao} = 1/\sqrt{c_{ao}}$, which are considered as given (one could propose a generative model for them too, but that would require further analysis).  
Then the probability density for $s$ is 
$$P(s|v,b,w) = \prod \frac{e^{-c_{ao}(s_{ao}-v_o-b_a)^2/2w}}{\sqrt{2\pi w/ c_{ao}}}
= (2\pi w)^{-N/2} e^{-\frac{1}{2w}\sum c_{ao}(s_{ao}-v_o-b_a)^2} \prod c_{ao}^{1/2}.$$

For prior probability density over the parameters $v,b,w$, we want to build in a degeneracy-breaking condition.  We used $\sum_a b_a = 0$ in our calculation, thus we take prior ``density''
$$P_{CWC}(v,b,w) = \frac{1}{Z_OZ_ALw}\delta(\sum_a b_a)$$
on the product of the balls (\ref{eq:ballO}) and $\sum_a b_a^2 \le N_A \sigma_A^2$ and interval $[w_{min},w_{max}]$, where $\delta$ is the delta function.
Here, $\sigma_A$ is an estimated upper bound for the standard deviation of the biases, and the normalisation is
$$Z_A = \frac{1}{\sqrt{N_A}} \frac{(\pi N_A \sigma_A^2)^{(N_A-1)/2}}{\Gamma((N_A+1)/2)}.$$
Note that the interpretation of $w$ is not the same as for SA, so one might choose a different prior for it.  For example, if the $\sigma_{ao}$ are fairly accurate values for the uncertainties in the scores then the prior for $w$ should be peaked around $w=1$, but if they are on an undetermined scale a truncated Jeffreys prior is sensible.  The only thing is that one might want to choose a different interval for it, but for application to IBA where the $c_{ao}=1$ or to CWC if the $c_{ao}$ are on a scale centred around 1, such as we have used to translate the quantitative high/medium/low confidence ratings, the same interval should be reasonable. 
Similarly, one might want to use a different value for $\sigma_O$ if one believes that the spread in values is more due to variation in assessor bias than true value, but in our case we think it reasonable to use the same $\sigma_O$.

Thus
$$P(s|v,b,w)P_{CWC}(v,b,w) = \frac{1}{Z_OZ_ALw} (2\pi w)^{-N/2} e^{-\frac{1}{2w}\sum c_{ao} (v_o+b_a-s_{ao})^2} \delta(\sum_a b_a) \prod \sqrt{c_{ao}}.$$
Again we assume the bulk of this lies in the product of balls and interval, so we approximate its integral by extending the domains to infinity.
Now
$$\sum c_{ao} (s_{ao}-v_o-b_a)^2 = h^T A h + R,$$
where $h$ is the vector with $N_O+N_A$ components, $h_o = (v_o-\bar{v}_o)$, $h_a = (b_a - \bar{b}_a)$, $(\bar{v},\bar{b})$ is any least squares fit to this model (without loss of generality satisfying the degeneracy-breaking condition), the residual $R$ is now $\sum c_{ao} (s_{ao}-\bar{v}_o-\bar{b}_a)^2$ (as in (\ref{eq:R})) and $A$ is the matrix with block form 
$$\left[\begin{array}{cc}
\mbox{diag}(C_o) & c^T \\
c & \mbox{diag}(C'_a) 
\end{array}
\right].$$

Choose one assessor, say $n$, and integrate over $b_n$.  This yields
$$\frac{1}{Z_OZ_ALw} (2\pi w)^{-N/2} e^{-\frac{1}{2w}(\tilde{h}^T \tilde{A} \tilde{h} + R)} \prod \sqrt{c_{ao}},$$
with $\tilde{h}$ being the remaining components of $h$ and
$$\tilde{A} = \left[\begin{array}{cc}
\mbox{diag}(C_o) & \tilde{c}^T \\
\tilde{c} & \mbox{diag}(C'_a) + C'_nE
\end{array} \right],$$
where $\tilde{c}_{ao} = c_{ao} - c_{no}$ and $E_{aa'} = 1$, restricted to $a,a' \ne n$, which takes into account that $b_n = - \sum_{a\ne n} b_a$.

Thus the integral over $\tilde{h}$ is 
$$\frac{1}{Z_OZ_ALw} (2\pi w)^{-\nu/2} e^{-R/2w}  \frac{\prod \sqrt{c_{ao}}}{\sqrt{\det{\tilde{A}}}},$$
where $\nu = N- N_O - N_A + 1$.

Finally, we integrate over $w$ to obtain
$$P(CWC|s) = \frac{1}{Z_OZ_AL} (\pi R)^{-\nu/2} \Gamma(\frac{\nu}{2}) \frac{\prod \sqrt{c_{ao}}}{\sqrt{\det{\tilde{A}}}},$$
and the log-evidence is
$$LE(CWC|s) = \log\Gamma(\frac{\nu}{2}) - \frac{\nu}{2}\log \pi R + \frac12 \sum \log c_{ao} - \frac12 \log \det{\tilde{A}} - \log Z_O - \log Z_A - \log L.$$

\subsection*{-- Incomplete Block Analysis}

The model for incomplete block analysis (IBA) is the same as for CWC but taking the confidences $c_{ao}=1$ for all the assessments.  Thus the log-evidence for IBA given the scores $s$ is
$$LE(IBA|s) = \log\Gamma(\frac{\nu}{2}) - \frac{\nu}{2}\log \pi R - \frac12 \log \det{\tilde{A}} - \log Z_O - \log Z_A - \log L,$$
with the appropriate changes to $R$ and $\tilde{A}$.

\section*{Appendix F: Potential Refinements to the method}

One could develop refinements to the basic model (\ref{eq:model1}).
For example, assessors might have not only an additive bias but also different scales, so for example
\begin{equation}
s_{ao} = \lambda_a v_o + b_a + \sigma_{ao} \eta_{ao} .
\label{eq:model2}
\end{equation}
Fitting $\lambda, v, b$ is more complicated, however, than just $v,b$.

An assessor may have a bias correlated with their confidence \cite{Song} or with some other feature like familiarity \cite{Fuchs}.
Assessors may like to give round-number scores\blue{ or the organisers of the panel may insist on them}.  \blue{Assessors} may have different scales for confidence,
so their confidences may need calibrating as well as their scores.  

Another problem is that often assessors are asked to assign scores in a fixed range $[A,B]$, e.g.~$1-10$.
Then any model for bias really ought to be nonlinear to respect the endpoints.  One way to treat this is to apply a nonlinear transformation to map a slightly larger interval $(a,b)$ onto $\R$, e.g.
\begin{equation}
 x \mapsto s = \frac{x-\frac12(a+b)}{(b-x)(x-a)}
 \end{equation}
 or
 \begin{equation}
 x \mapsto s = \log\frac{b-x}{x-a},
 \end{equation}
apply our method to the transformed scores, scaling the confidences by the inverse square of the derivative of the transformation, and then apply the inverse transformation to the ``true" values.
On the other hand, it may be inadvisable to specify a fixed range because it requires an assessor to have knowledge of the range of the objects before starting scoring.  Thus one could propose asking assessors to use any real numbers and then use equation~(\ref{eq:model2}) to extract true values $v$.
A simpler strategy that might work nearly as well is to allow assessors to use any positive numbers but then to take logarithms and fit equation~(\ref{eq:model1}) to the log-scores.  The assessor biases would then be like logarithms of exchange rates.  The confidences would need translating appropriately too.

One issue with our method is that the effect of an assessor who assesses only one object is only to determine their own bias, apart from an overall shift along the null vector $(v,b)=(1,-1)$ for the rest.  To rectify this one could incorporate a prior probability distribution for the biases (indeed, this was done by \cite{PB} in the form of a regulariser).

An interesting future project is to design the graph $\Gamma$ optimally, given advance guesses of confidences and constraints \blue{(like conflicts of interest)} or costs for the number of assessments per assessor.  ``Optimality'' would mean to achieve maximum precision or robustness of the resulting values.  For instance, in each case of Figure~\ref{fig1:graphs}, each assessor has the same amount of work and each object receives the same amount of attention, but (a) achieves full connectivity with a resulting value for $\mu_2$ of $1-\sqrt{1/3} \approx 0.42$, whereas (b) achieves moderate connectivity and a smaller value of $\mu_2 = 1-\sqrt{2/3} \approx 0.18$, and (c) is not even connected and has $\mu_2 = 0$.

\end{document}